\documentclass[secnumarabic,aps,pra,amsfonts,twocolumn,%
showkeys,floatfix%
]{revtex4}

\usepackage{bm,longtable,ams,amsfonts}
\usepackage{graphicx}
\newcommand{\gl}[1]{(\ref{#1})}

\begin{document}

\title{Stability of the superconducting state in YBa$_2$Cu$_3$O$_7$} 
\author{Ekkehard Kr\"uger}
\affiliation{Max-Planck-Institut f\"ur Metallforschung, D-70506 Stuttgart,
  Germany}
%
%
\begin{abstract}
  The nonadiabatic Heisenberg model (NHM) proposed as an extension of the
  Heisenberg model makes a contribution to the {\em eigenstate} problem of
  superconductivity.  The Hamiltonian $H^n$ derived within this
  group-theoretical model has superconducting eigenstates if {\em and only if}\ 
  the considered material possesses a narrow, roughly half-filled
  ``superconducting'' energy band of special symmetry in its band structure.
  This paper shows that the high-temperature superconductor YBa$_2$Cu$_3$O$_7$
  possesses such a superconducting band. This new result together with previous
  observations about other superconductors and non-superconductors corroborates
  the theoretical evidence within the NHM that stable superconducting states
  are connected with superconducting bands. It is proposed that the type of
  superconductivity, i.e., whether the material is a conventional low-$T_c$ or
  a high-$T_c$ superconductor, is determined by the energetically lowest boson
  excitations that carry the crystal spin $1\cdot\hbar$ and are sufficiently
  stable to transport this crystal spin-angular momentum through the crystal.
  This mechanism provides the electron-phonon mechanism that enters the BCS
  theory in conventional superconductors.
\end{abstract}
\maketitle

\section{Introduction}
To date, it is not possible to solve the Schr\"odinger equation
\begin{equation}
  \label{eq:4}
H\psi = E\psi    
\end{equation}
for superconducting or non-superconducting materials. Hence, the traditional
theory of both conventional low-$T_c$ and high-$T_c$ superconductivity has to
postulate the existence of superconducting eigenstates $\psi$. The results of
the traditional theory have a physical meaning only if superconducting
eigenstates exist for the considered material.

In former papers~\cite{es,enhm} the author proposed an extension of the
Heisenberg model of magnetism~\cite{hei}, called nonadiabatic Heisenberg model
(NHM). This group-theoretical model goes beyond the adiabatic approximation and
gives a contribution to the eigenstate problems of both magnetism and
superconductivity.  Within the NHM, superconducting eigenstates are connected
with the existence of a narrow, roughly half-filled ``superconducting band'' in
the band structure of the considered material.  In such a superconducting band
the related nonadiabatic Hamiltonian $H^n$ derived within the NHM undoubtedly
has superconducting eigenstates because $H^n$ acts in a special part of the
Hilbert space which may be called ``Cooper space''~\cite{josn}. This space
represents a nonadiabatic system in which {\em constraining forces} are
effective in a way familiar from the classical mechanics. Below a transition
temperature $T_c$, these constraining forces reduce the degrees of freedom of
the electron system by forcing the electrons to form pairs that are invariant
under time inversion, i.e., by forcing the electrons to form Cooper pairs that
possess only half the degrees of freedom of unpaired electrons.  Since this
pair formation is mediated by boson excitations, the traditional theory of
superconductivity provides within the Cooper space the quantitative methods to
calculate $T_c$ and other parameters of the superconducting state.

In materials that do not possess a superconducting band, on the other hand,
constraining forces that halve the degrees of freedom of the electrons do
not exist. In the classical mechanics, however, any reduction of the degrees of
freedom of any system of particles is caused by constraining forces. Hence, it
cannot be excluded that also in quantum mechanical systems any reduction of the
electronic degrees of freedom is produced by constraining forces since quantum
particles behave in some respects similar to classical particles. This
comparison of the quantum system with a classical system suggests that the
constraining forces established within the NHM in superconducting bands {\em
  are required} for the formation of Cooper pairs, i.e., they are required for
the Hamiltonian $H$ to possess superconducting eigenstates. Consequently, there
is evidence that materials which do not possess a narrow, roughly half-filled
superconducting band, do not become superconducting even if the traditional
theory of superconductivity predicts a stable superconducting state.

Superconducting bands have already been identified in the band structures of a
large number of elemental superconductors \cite{es2} and of the
high-temperature superconductor La$_2$CuO$_4$ \cite{la2cuo4}.  The theoretical
evidence that superconducting bands are required for superconducting
eigenstates to exist is corroborated by the fact that superconducting bands
cannot be found in those elemental metals (such as Li, Na, K, Rb, Cs, Ca Cu,
Ag, and Au) which do not become superconducting \cite{es2}. An investigation
into the band structures of the transition metals in terms of superconducting
bands straightforwardly leads to the Matthias rule \cite{josm}.

The aim of the present paper is to investigate whether or not the
superconducting state of the high-$T_c$ superconductor YBa$_2$Cu$_3$O$_7$ found
by C. W. Chu, M. K. Wu, and co-workers~\cite{wu} is also connected with the
existence of a superconducting band. Before in Sec.~\ref{sec:sband} the
existence of this band shall be established, in the following
Sec.~\ref{sec:nhm} the NHM is shortly described, in Sec.~\ref{sec:mstate} it is
shown that the antiferromagnetic structure existing in YBa$_2$Cu$_3$O$_6$ is
not stable in YBa$_2$Cu$_3$O$_7$, and Sec.~\ref{sec:sbands} gives the
definition of superconducting bands and a short characterization of the
mechanism of Cooper pair formation in these bands.

\section{Nonadiabatic Heisenberg model}
\label{sec:nhm}
In the framework of the NHM, the electrons in narrow, roughly half-filled bands
may lower their Coulomb energy by occupying an atomiclike state as proposed by
Mott \cite{mott} and Hubbard \cite{hubbard}: as long as possible the electrons
occupy localized states and perform their band motion by hopping from one atom
to another. The localized states are represented by localized functions
depending on an additional coordinate\ $\vec q$~related to the nonadiabatic
motion of the center of mass of the localized state.  The introduction of this
new coordinate allows a consistent specification of the atomiclike motion that
has a {\em lower energy} than a purely bandlike motion. The nonadiabatic
localized functions {\em must} be adapted to the symmetry of the crystal so
that the nonadiabatic Hamiltonian $H^n$ has the correct commutation properties.

The nonadiabatic localized functions may be approximated by the best localized
(spin-dependent) Wannier functions when the nonadiabatic motion of the centers
of mass again is disregarded. These ``adiabatic'' and our nonadiabatic
functions will not differ so strongly that their symmetry is altered at the
transition from the adiabatic to the nonadiabatic system.  Hence, suitable
nonadiabatic localized states allowing an atomiclike motion of the electrons
exist if and only if best localized Wannier functions exist which are adapted
to the symmetry of the crystal.

\section{Instability of the magnetic ordered state in
  $\text{YBa}_2\text{Cu}_3\text{O}_7$} 
\label{sec:mstate}


\begin{figure}
\begin{center}
(a)\vspace{.5cm}

\includegraphics[width=.2\textwidth,angle=0]{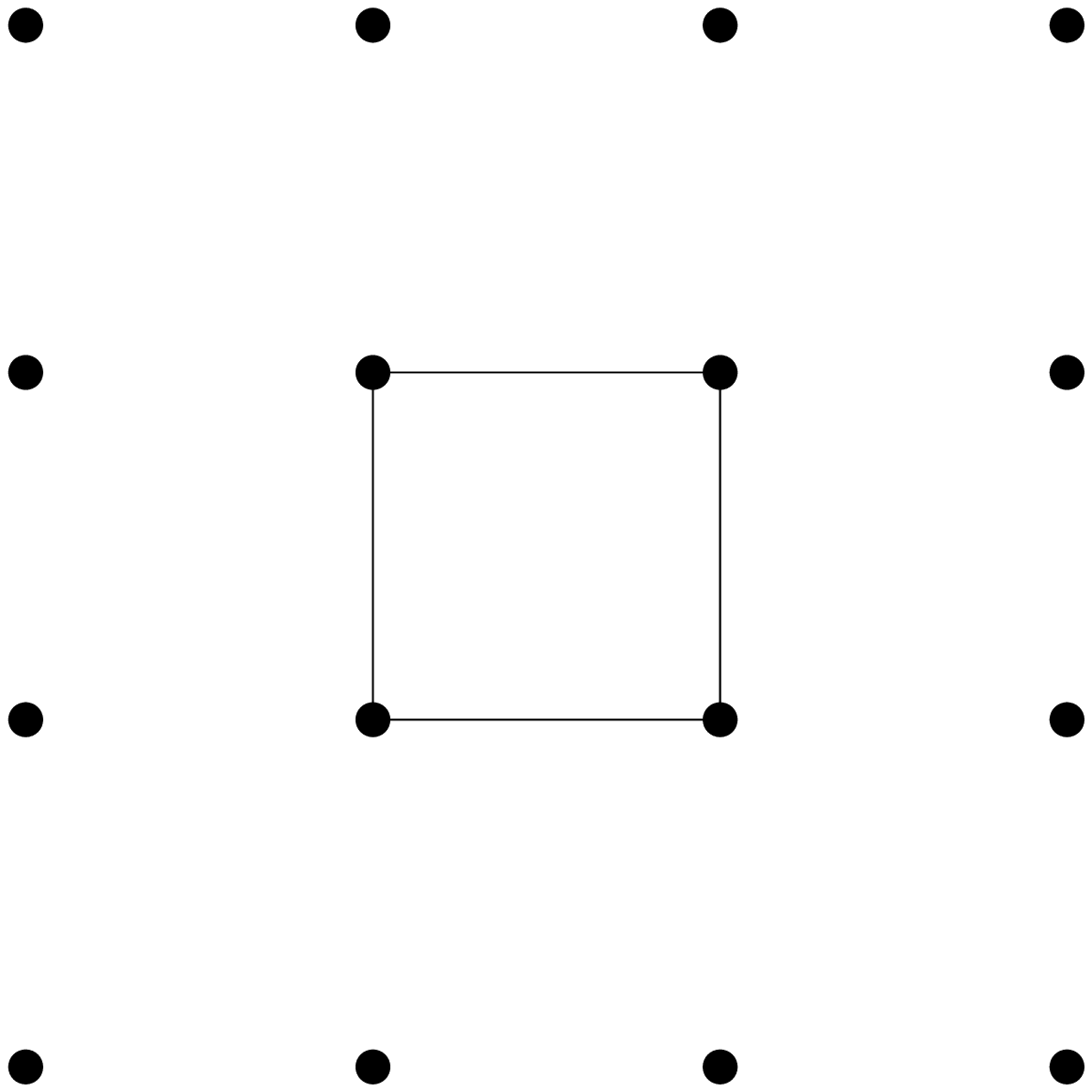}%
\vspace{1cm}

(b)\vspace{.5cm}

\includegraphics[width=.2\textwidth,angle=0]{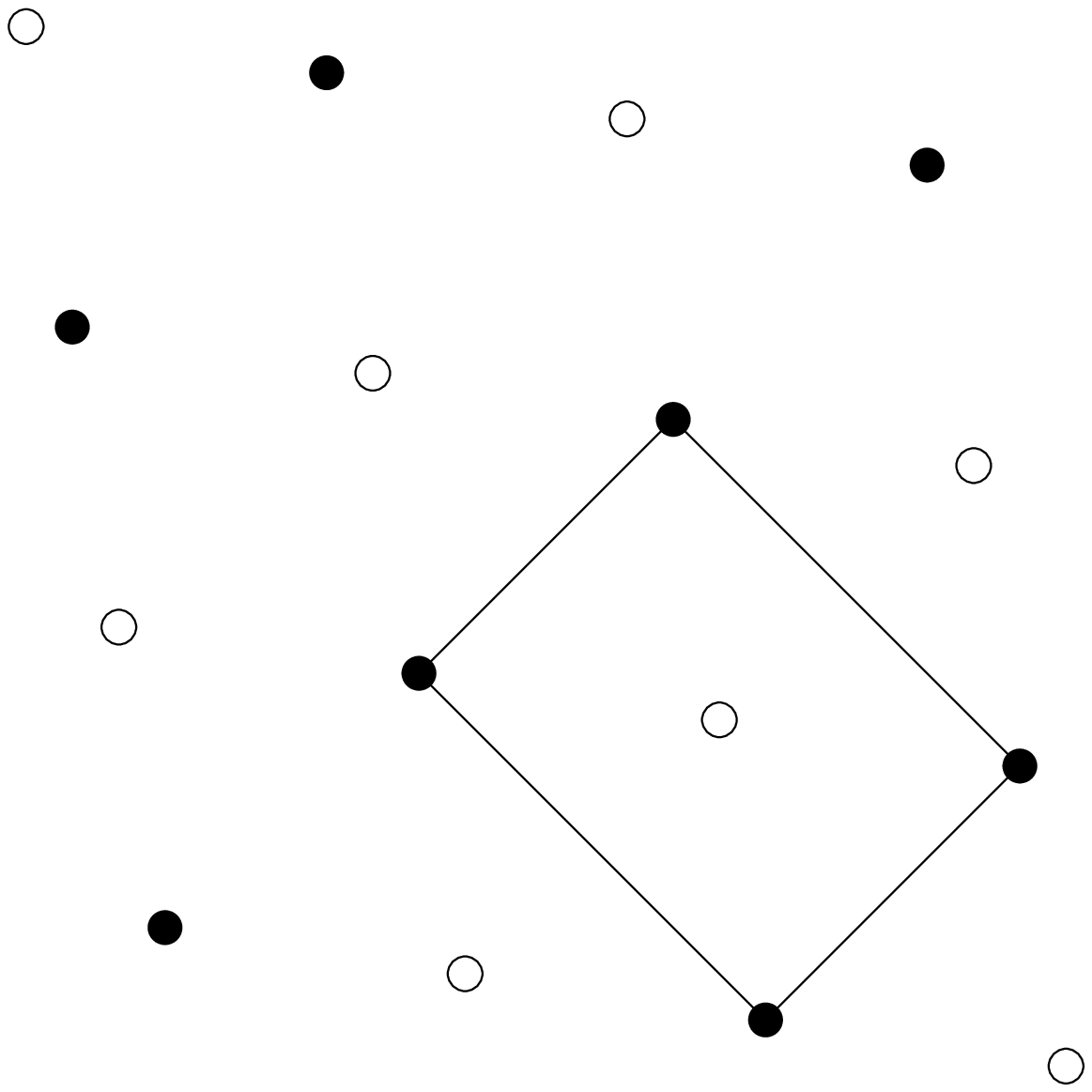}%
\vspace{1cm}

(c)\vspace{.5cm}

\includegraphics[width=.2\textwidth,angle=0]{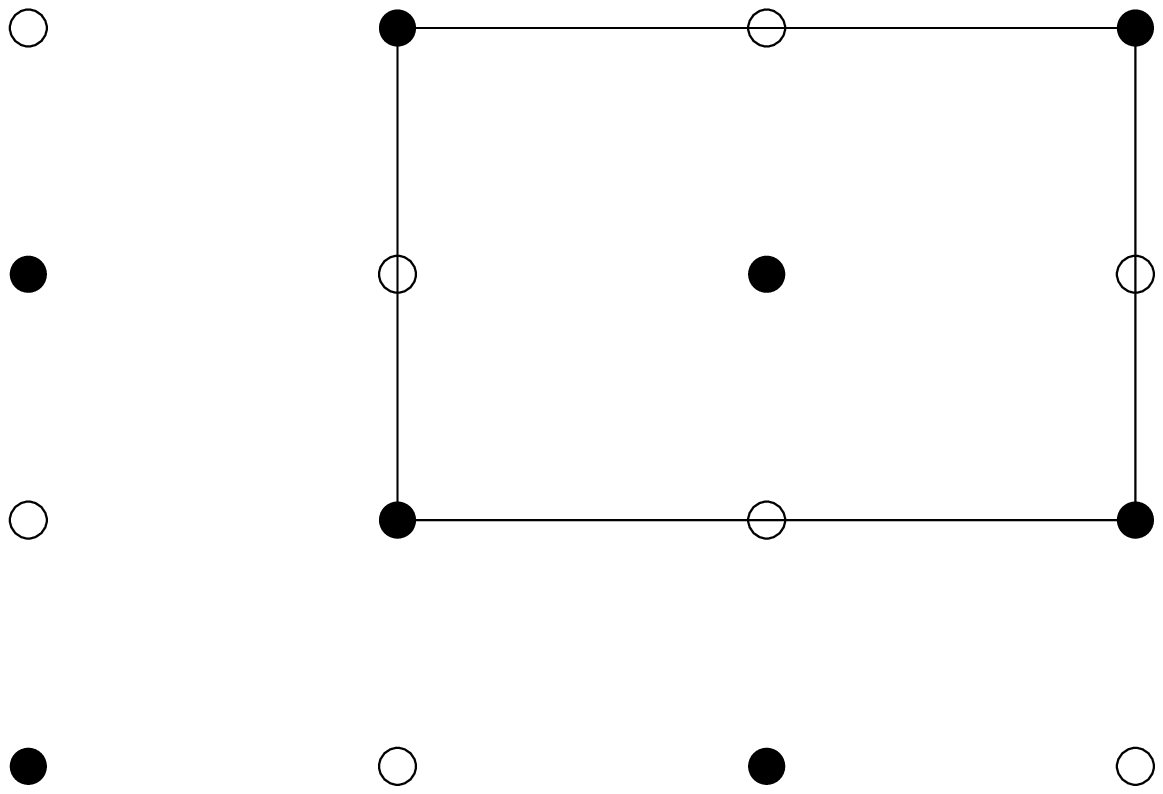}%
\end{center}
\caption{
  Qualitative orthorhombic distortions of the Cu sites in the A (or C) layers
  of YBa$_2$Cu$_3$O$_6$ and YBa$_2$Cu$_3$O$_7$, respectively.  (a) The
  undistorted Cu sites in the tetragonal lattice of paramagnetic
  YBa$_2$Cu$_3$O$_6$. (b) The distortion with the orthorhombic primitive
  Bravais lattice $\Gamma_o$ produced by the magnetic structure in
  YBa$_2$Cu$_3$O$_6$. Solid and open circles represent Cu sites with mutually
  antiparallel magnetic moments. (c) The distortion produced by the additional
  O atom in YBa$_2$Cu$_3$O$_7$. Solid and open circles represent Cu sites with
  mutually antiparallel magnetic moments in a structure corresponding to the
  magnetic structure in YBa$_2$Cu$_3$O$_6$. This magnetic structure, if it
  exists, possesses the orthorhombic base-centered Bravais lattice
  $\Gamma^b_o$.
\label{fig:Gitter}
}
\end{figure}


The antiferromagnetic structure in YBa$_2$Cu$_3$O$_6$ as well as the additional
O atom in YBa$_2$Cu$_3$O$_7$ leads to an orthorhombic distortion of the
paramagnetic tetragonal crystal. Hence, one could believe that the additional O
atom stabilizes the magnetic structure observed in YBa$_2$Cu$_3$O$_6$.
However, this it not true because the orthorhombic distortions are basically
different.  While the magnetic phase in YBa$_2$Cu$_3$O$_6$ has the orthorhombic
primitive Bravais lattice $\Gamma_o$, the corresponding magnetic structure in
YBa$_2$Cu$_3$O$_7$ would possess the orthorhombic base-centered lattice
$\Gamma^b_o$, see Fig.~\ref{fig:Gitter}. In YBa$_2$Cu$_3$O$_7$ the magnetic
group would have the form
\begin{equation}
  \label{eq:2}
 M = H + \{K|\textstyle\frac{1}{2}\frac{1}{2}0\}H, 
\end{equation}
where $K$ stands for the operator of time inversion and $H$ denotes a space
group with the Bravais lattice $\Gamma^b_o$. 

Within the NHM, both a magnetic state $|m\rangle$ and the time-inverted
state
\[
\overline{|m}\rangle = K|m\rangle   
\]
are {\em eigenstates} of a Hamiltonian commuting with the operator $K$ of time
inversion. Hence, a {\em stable} magnetic state $|m\rangle$ complies with two
conditions: 
\begin{itemize}
\item 
$|m\rangle$ is basis function of a one-dimensional corepresentation
of $M$;
\item $|m\rangle$ and the time-inverted state $\overline{|m}\rangle$ are
  basis functions of a two-dimensional {\em irreducible}\/ corepresentation
  of the gray magnetic group
\begin{equation}
\overline{M} = M + KM,
\label{graymgroup}
\end{equation}
\end{itemize}
see Sec.~III.C of Ref.\ \cite{ea}.

As in the forgoing paper~\cite{ybacuo6} we have to look for
space groups $H$ possessing at least one one-dimensional single-valued
representation $R$ following
\begin{enumerate}
\item 
case (a) with respect to the
magnetic group $H + \{K|\frac{1}{2}\frac{1}{2}0\}H$ and 
\item 
case (c) with
respect to the magnetic group \mbox{$H + KH$.}
\end{enumerate}
The cases (a) and (c) are defined in Eqs.\ (7.3.45) and (7.3.47),
respectively, in the textbook of Bradley and Cracknell \cite{bc}. The
irreducible corepresentation derived from $R$ stays one-dimensional in case
(a) and becomes two-dimensional in case (c).

Among all the space groups with the Bravais lattice $\Gamma^b_o$ there are only
the two groups $H_1 = \Gamma^b_oC^{12}_{2v}$ (36) and $H_2 =
\Gamma^b_oC^{13}_{2v}$ (37) possessing one-dimensional representations
following case (c) with respect to \mbox{$H_i + KH_i\ (i = 1,2)$.} (The number
in parentheses is the international number of the space group.) That means,
only these space groups have non-real one-dimensional representations, see
Table 5.7 of Ref.~\cite{bc}. However, applying Eq. 7.3.51 of Ref.~\cite{bc} to
the non-real representations of $H_1$ or $H_2$, it turns out that all of them
also follow case (c) with respect to $H_i + \{K|\frac{1}{2}\frac{1}{2}0\}H_i\ 
(i = 1,2)$. Hence, a magnetic structure with the Bravais lattice
$\Gamma^b_o$ and with the anti-unitary operation $\{K|\frac{1}{2}\frac{1}{2}0\}$
is not stable in YBa$_2$Cu$_3$O$_7$. This material does not possess a magnetic
structure corresponding to the magnetic structure in YBa$_2$Cu$_3$O$_6$.
However, it cannot be excluded that other, more complex magnetic structures
might be stable in YBa$_2$Cu$_3$O$_7$.


 \begin{figure*}[t]
  \includegraphics[width=.65\textwidth,angle=-90]{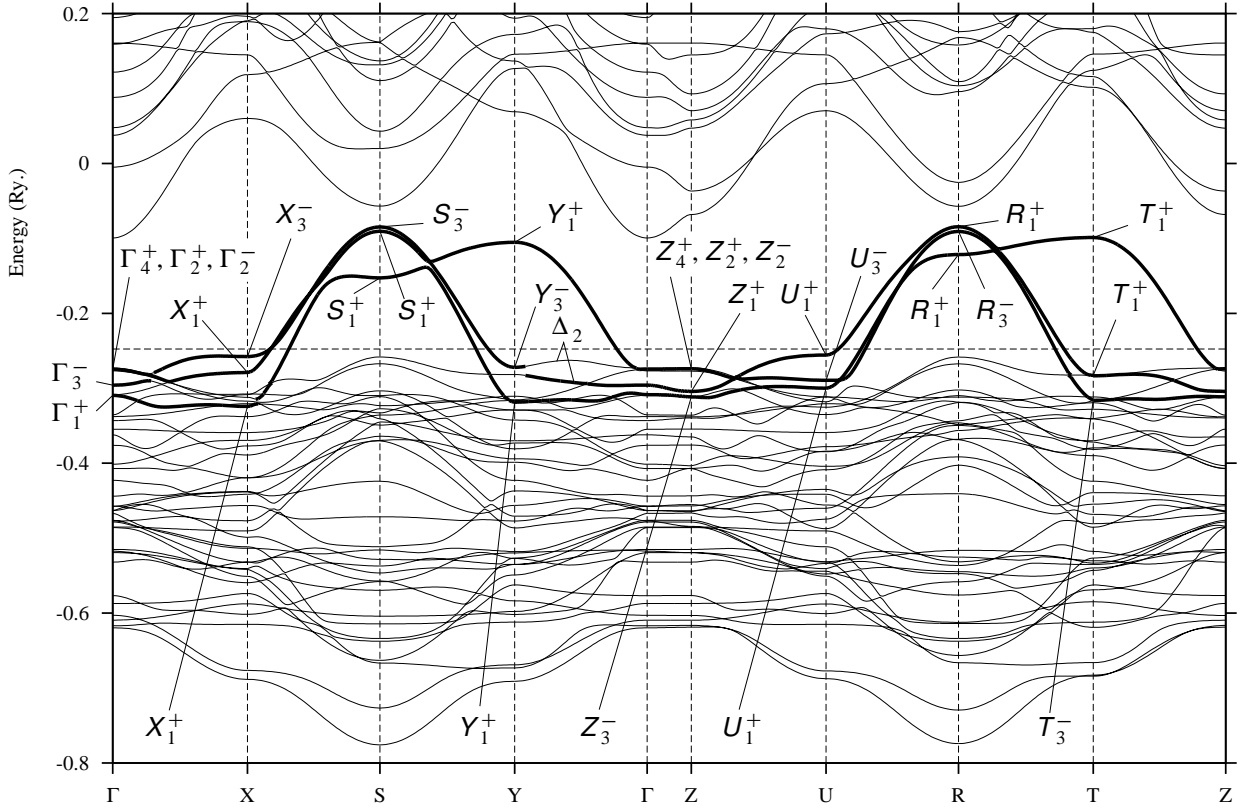}
 \caption{
   Band structure of orthorhombic YBa$_2$Cu$_3$O$_7$ as calculated by Ove
   Jepsen \cite{jepsen}, with symmetry labels determined by the author. The
   symmetry labels can be identified from Table~\ref{tab:darst}. The bold line
   shows the superconducting band consisting of three branches.
  \label{fig:bandstr}
 }
 \end{figure*}


\section{Superconducting bands}
\label{sec:sbands}
In YBa$_2$Cu$_3$O$_6$ the Bloch functions of the half-filled
``antiferromagnetic band'' can be unitarily transformed into best localized
Wannier functions situated on the Cu sites and adapted to the symmetry of the
magnetic structure existing in this material. Hence, the electrons may lower
their Coulomb energy by occupying an atomiclike state represented by these
Wannier functions \cite{ybacuo6}. In YBa$_2$Cu$_3$O$_7$, on the other hand, the
electrons cannot lower their Coulomb energy in the same way because the
magnetic structure existing in YBa$_2$Cu$_3$O$_6$ is not stable in
YBa$_2$Cu$_3$O$_7$.

In YBa$_2$Cu$_3$O$_7$ the electrons have another way to occupy the
energetically favorable atomiclike state. As I shall show in the following
Sec.~\ref{sec:sband}, this material possesses a narrow, roughly half-filled
``superconducting band'' \cite{es,josn,enhm} in its band structure, see
Fig.~\ref{fig:bandstr}.  An energy band of YBa$_2$Cu$_3$O$_7$ is called
``superconducting band'' when the Bloch functions of this band can be unitarily
transformed into {\em spin-dependent} Wannier functions which are
\begin{itemize}
\item centered on the Cu sites;
\item symmetry-adapted to the space group $H = \Gamma_oD^{1}_{2h}$ of this
  material; and
\item best localized.
\end{itemize}
In this context I assume the localized states to have the same positions as
they have in the antiferromagnetic state of YBa$_2$Cu$_3$O$_6$ \cite{ybacuo6}.
A superconducting band in YBa$_2$Cu$_3$O$_7$ consists of three branches because
there are three Cu atoms in the unit cell.  Since YBa$_2$Cu$_3$O$_7$ possesses
a narrow, roughly half-filled superconducting band, the electrons in this
material may lower their Coulomb energy by occupying an atomiclike state
represented by spin-dependent Wannier functions. In this section I shall
summarize the ideas leading to the suggestion that such an atomiclike state is
superconducting at sufficiently low temperatures. A detailed theoretical
substantiation is given in Ref.~\cite{josn}, the physical background is
examined ´more closely in Refs.~\cite{josi} and~\cite{ekm}.


\begin{table}[!]
\caption{Character tables of the irreducible representations $R^{\pm}_i$ of the
  space group $Pmmm = \Gamma_oD^{1}_{2h}$ (47) of YBa$_2$Cu$_3$O$_7$. 
\label{tab:darst}}
\begin{center}
\em{(a)\quad Single-valued representations}
\end{center}
\begin{tabular}[t]{ccccccccc}
 & $E$ & $C_{2z}$ & $C_{2y}$ & $C_{2x}$ & $I$ & $\sigma_z$ & $\sigma_y$ & $\sigma_x$\\
\hline
$R^+_1$ & 1 & 1 & 1 & 1 & 1 & 1 & 1 & 1\\
$R^+_2$ & 1 & -1 & 1 & -1 & 1 & -1 & 1 & -1\\
$R^+_3$ & 1 & 1 & -1 & -1 & 1 & 1 & -1 & -1\\
$R^+_4$ & 1 & -1 & -1 & 1 & 1 & -1 & -1 & 1\\
$R^-_1$ & 1 & 1 & 1 & 1 & -1 & -1 & -1 & -1\\
$R^-_2$ & 1 & -1 & 1 & -1 & -1 & 1 & -1 & 1\\
$R^-_3$ & 1 & 1 & -1 & -1 & -1 & -1 & 1 & 1\\
$R^-_4$ & 1 & -1 & -1 & 1 & -1 & 1 & 1 & -1\\
\hline\\
\end{tabular}
\begin{center}
\em{(b)\quad Double-valued representations}
\end{center}
\begin{tabular}[t]{ccccccccccc}
 & $$ & $$ & $\overline{C}_{2z}$ & $\overline{C}_{2y}$ & $\overline{C}_{2x}$ &
 $$ & $$ & 
 $\overline{\sigma}_z$ & $\overline{\sigma}_y$ & $\overline{\sigma}_x$\\ 
 & $E$ & $\overline{E}$ & $C_{2z}$ & $C_{2y}$ & $C_{2x}$ & $I$ &
 $\overline{I}$ & $\sigma_z$ & $\sigma_y$ & $\sigma_x$\\ 
\hline
$R^+_5$ & 2 & -2 & 0 & 0 & 0 & 2 & -2 & 0 & 0 & 0\\
$R^-_5$ & 2 & -2 & 0 & 0 & 0 & -2 & 2 & 0 & 0 & 0\\
\hline\\
\end{tabular}
\footnotetext{
\ \\
Notes to Table~\ref{tab:darst}
\begin{enumerate}
\item The two tables (a) and (b) are common to all the points of symmetry
  $\Gamma (000)$, $Y (\overline{\frac{1}{2}}00)$,
  $S(\overline{\frac{1}{2}}\frac{1}{2}0)$, $X (0\frac{1}{2}0)$,
  $Z(00\frac{1}{2})$, $T(\overline{\frac{1}{2}}0\frac{1}{2})$,
  $R(\overline{\frac{1}{2}}\frac{1}{2}\frac{1}{2})$, and $U
  (0\frac{1}{2}\frac{1}{2})$ in the Brillouin zone for $\Gamma_o$.
\item The letter $R$ has to be replaced by the letter denoting the relevant
  point of symmetry.
\item The tables are determined from Tables 5.7 and 6.13 in the textbook of
  Bradley and Cracknell \protect\cite{bc}.
\end{enumerate}
}
\end{table}


Within the adiabatic approximation, an atomiclike motion with localized
functions depending on the electron spin does not conserve the spin angular
momentum. In the (real) nonadiabatic system, however, the electrons may
interchange spin angular momenta with the lattice of the atomic cores. In the
framework of the NHM, an atomiclike motion with localized functions depending
on the electron spin leads to a special spin-boson interaction in the
nonadiabatic system: at any electronic scattering process two crystal-spin-1
bosons are excited or absorbed.

At low temperature, this spin-boson interaction has a striking feature
distinguishing it from any usual electron-phonon or electron-boson interaction:
it {\em constrains} the electrons in a special way to form electron pairs
invariant under time inversion because the conservation of spin angular
momentum would be violated in any unpaired state. This mechanism of pair
formation shows resemblances, but also an important difference when compared
with the familiar mechanism of Cooper pair formation presented within the
Bardeen-Cooper-Schrieffer (BCS) theory~\cite{bcs}. The difference between both
mechanisms can be described in terms of constraining forces that halve the
degrees of freedom of the electrons~\cite{josn}, or in terms of
``spring-mounted'' Cooper pairs~\cite{josi}.

Electron pairs invariant under time inversion cause superconductivity. The
superconducting transition temperature may be approximately calculated by the
familiar BCS formula
\begin{equation}
T_{c} = 1.14\cdot\theta\cdot e^{-1/N(E_F)V}
\label{bcs}
\end{equation}
where now $N(E_{F})$, $V$ and $\theta$ are the density of states of the
electrons of the superconducting band at the Fermi level, the effective
spin-phonon interaction, and the Debye temperature, respectively.  The Debye
temperature $\theta$ now is related to the spectrum of the energetically lowest
excitations of the crystal possessing the crystal spin $1\cdot\hbar$ and being
sufficiently stable to transport it through the crystal, see also
Sec.~\ref{sec:disc}.

\section{Superconducting band in $\text{YBa}_2\text{Cu}_3\text{O}_7$}
\label{sec:sband}

In this section I show that the energy band denoted in Fig.~\ref{fig:bandstr}
by the bold line is a superconducting band with weakly spin-dependent Wannier
functions. It is labeled by the representations
\begin{equation}
\label{eq:1}
\begin{array}{l}
\underline{\Gamma^+_4},\Gamma^+_2,\Gamma^-_2,\underline{\Gamma^-_3},
\underline{\Gamma^+_1};\\
X^-_3, X^+_1, X^+_1;\\ 
S^-_3, S^+_1, S^+_1; \\
Y^+_1, Y^-_3, Y^+_1; \\ 
\underline{Z^+_4},Z^+_2,Z^-_2, \underline{Z^+_1}, \underline{Z^-_3};\\
U^+_1, U^-_3, U^+_1; \\
R^+_1, R^-_3, R^+_1; \\
T^+_1, T^+_1, T^-_3. \\   
\end{array}
\end{equation}


\begin{table}
\caption{
Single- and double-valued representations of all the energy bands in
YBa$_2$Cu$_3$O$_7$ with symmetry-adapted and optimally 
localized (spin-dependent) Wannier functions centered at the Cu atoms. 
\label{tab:atomwf}}
\begin{center}
  {\em (a)\quad Single-valued representations} 
\end{center}
\begin{tabular}[t]{ccc}
\hline
Band 1 & & 2$R^+_1$ + $R^-_3$ \\
Band 2 & & 2$R^+_2$ + $R^-_4$ \\
Band 3 & & 2$R^+_3$ + $R^-_1$ \\
Band 4 & & 2$R^+_4$ + $R^-_2$ \\
Band 5 & & $R^+_3$ + 2$R^-_1$ \\
Band 6 & & $R^+_4$ + 2$R^-_2$ \\
Band 7 & & $R^+_1$ + 2$R^-_3$ \\
Band 8 & & $R^+_2$ + 2$R^-_4$ \\
\hline\\
\end{tabular}
\begin{center}
  {\em (b)\quad Double-valued representations}
\end{center}
\begin{tabular}[t]{ccc}
\hline
Band 1 & & 2$R^+_5$ + $R^-_5$ \\
Band 2 & & $R^+_5$ + 2$R^-_5$ \\
\hline\\
\end{tabular}
\footnotetext{
Notes to Table~\ref{tab:atomwf}
\begin{enumerate}
\item The two tables (a) and (b) are common to all the points of symmetry
  $\Gamma$, $Y$, $S$, $X$, $Z$, $T$, $R$, and $U$ in the Brillouin zone for
  $\Gamma_o$.
\item The letter $R$ has to be replaced by the letter denoting the relevant
  point of symmetry.
\item The two bands with the double-valued representations (b) form
  superconducting bands.
\item The representations can be identified from Table~\ref{tab:darst}.
\item Each row defines one band consisting of three branches, because there
  are three Cu atoms in the unit cell.
\item The bands are determined by Eq.~(B7) of Ref.~\protect\cite{la2cuo4}. 
\item Assume a band of the symmetry in any row of this table to exist in the
  band structure of a given material with the space group $\Gamma_oD^{1}_{2h}$.
  Then the Bloch functions of this band can be unitarily transformed into
  Wannier functions that are
\begin{itemize}
\item as well localized as possible; 
\item centered at the Cu atoms; and
\item symmetry-adapted to the space group $\Gamma_oD^{1}_{2h}$. 
\end{itemize}
These Wannier function are usual (spin-independent) Wannier function if the
considered band is characterized by the single-valued representations (a). They
are spin-dependent if the band is characterized by the double-valued
representations (b).
\end{enumerate}
}
\end{table}


Table~\ref{tab:atomwf} (a) lists all the (eight) bands in YBa$_2$Cu$_3$O$_7$
whose Bloch functions can be unitarily transformed into symmetry-adapted and
best localized Wannier functions situated on the Cu sites. Each band consists
of three branches because there are three Cu atoms in the unit cell. While the
representations~\gl{eq:1} coincide with the representations of band 1 in
Table~\ref{tab:atomwf} (a) at points $X, S, Y, U, R,$ and $T$, the
representations at points $\Gamma$ and $Z$ are slightly different: band 1 in
Table~\ref{tab:atomwf} (a) contains $\Gamma^+_1$ as well as $Z^+_1$ twice, but
these representations are found only once among the representations~\gl{eq:1}.
Hence, we {\em cannot} represent the Bloch functions of this band by
symmetry-adapted and best localized Wannier functions situated on the Cu sites.

The situation is changed when we replace the single-valued representations
$R^{\pm}_i$ by the corresponding {\em double-valued} representations
$R^{\pm}_i\times D_{1/2}$. From Table~\ref{tab:darst} (b) we can derive
that 
\begin{equation}
  \label{eq:3}
  R^{\pm}_i\times D_{1/2} = R^{\pm}_5
\end{equation}
holds for any representation $R^{\pm}_i$ in Table~\ref{tab:darst} (a).
$D_{1/2}$ denotes the two-dimensional double-valued representation of the
three-dimensional rotation group $O(3)$ given, e.g., in Table 6.1 of
Ref.~\cite{bc}. Table~\ref{tab:atomwf} (b) lists all the superconducting bands
in YBa$_2$Cu$_3$O$_7$. The energy band characterized by the
representations~\gl{eq:1} now becomes identical with band 1 in
Table~\ref{tab:atomwf} (b): The Bloch functions of the band~\gl{eq:1} can be
unitarily transformed into symmetry-adapted and best localized {\em
  spin-dependent} Wannier functions situated on the Cu sites. At points
$\Gamma$ and $Z$ we may use the Bloch functions with the underlined
representations~\gl{eq:1}.

The Wannier functions are {\em weakly}\ spin-dependent since they do not
strongly differ from the spin-{\em in}dependent Wannier function belonging to
band 1 in Table~\ref{tab:atomwf} (a). It is only one of the two representations
$2\Gamma^+_1$ and $2Z^+_1$, respectively, of band 1 in Table~\ref{tab:atomwf}
(a) which does not belong to the superconducting band~\gl{eq:1}.

\section{Discussion}
\label{sec:disc}
This paper shows that the high-temperature superconductor YBa$_2$Cu$_3$O$_7$
possesses a narrow, roughly half-filled ``superconducting band'' in its band
structure, see Fig.~\ref{fig:bandstr}, which enables the electrons to occupy
the energetically favorable atomiclike state represented by {\em
  spin-dependent} Wannier functions.  Within the NHM, the spin-dependence of
the localized functions has an important consequence: at zero temperature, the
electrons {\em must} form pairs invariant under time inversion because the
conservation of spin angular momentum would be violated in any unpaired state.
Hence, at zero temperature, the atomiclike state in a superconducting band is
superconducting of necessity. The related nonadiabatic Hamiltonian $H^n$
clearly has superconducting eigenstates.

This result suggests that the superconducting state in YBa$_2$Cu$_3$O$_7$ is
stabilized by the {\em constraining forces} that are effective in narrow
superconducting bands. Both magnetism and superconductivity have the same
physical origin: they exist because the electrons at the Fermi level tend to
occupy the energetically favorable {\em atomiclike} state. The special symmetry
properties of the related Wannier function determine whether the material
becomes magnetic or superconducting (or has a property not yet considered).
From this angle, superconductivity might also be called ``$\vec k$-space
magnetism'' \cite{ekm}.

The superconducting transition temperature is approximately determined by the
BCS equation~\gl{bcs} with the Debye temperature $\theta$ being related to the
spectrum of the {\em energetically lowest} boson excitations of the crystal
that possess the crystal spin $1\cdot\hbar$ and are sufficiently stable to
transport it through the crystal.  These ``crystal-spin-1'' bosons are
localized excitations $|\vec T, l\rangle$ (with $l = -1, 0, +1$ labeling the
three directions of the crystal spin and $\vec T$ denoting a lattice point) of
well-defined symmetry \cite{es,ehtc} which propagate as Bloch waves (with the
crystal momentum $\hbar\vec k$) through the crystal.

The $|\vec T, l\rangle$ are generated during spin-flip processes in the
superconducting band and must carry off the surplus spin angular momenta
generated at these processes. This spin-phonon mechanism suggests that the
$|\vec T, l\rangle$ are coupled modes of lattice vibrations and vibrations of
the core electrons against the atomic lattice: In a first step the atomiclike
electrons in the superconducting band transmit their angular momenta to the
core electrons by generating a plasmon-like vibration of the core electrons
against the atoms. In a second step, these plasmon-like excitations generate
phonon-like vibrations of lower energy if phonons of the appropriate symmetry
are sufficiently stable.

Thus, it is proposed that the $|\vec T, l\rangle$ are coupled phonon-plasmon
modes which determine the type of superconductivity.  They have dominant phonon
character in the isotropic lattices of the transition elements and, hence,
confirm the electron-phonon mechanism that enters the BCS theory in these
materials.  However, phonon-like excitations are not able to transport
crystal-spin angular-momenta within the two-dimensional copper-oxygen layers of
YBa$_2$Cu$_3$O$_7$, see, for preliminary ideas to this problem,
Ref.~\cite{ehtc}.  Within the two-dimensional layers, the $|\vec T, l\rangle$
necessarily are energetically higher lying excitations of dominant plasmon
character leading in the BCS equation~\gl{bcs} to a higher Debye temperature
$\theta$ and, hence, to a higher $T_{c}$.  This interpretation is corroborated
by the experimental evidence for two-dimensional superconductivity in the
CuO$_2$ atomic plains~\cite{prl_ling,harshman,bickers,monthoux,chakravarty}.

In the band structures of the high-temperature superconductors
YBa$_2$Cu$_3$O$_7$ (this paper) and La$_2$CuO$_4$~\cite{la2cuo4} I discovered
an absolutely new feature of the superconducting bands: the related Wannier
functions are {\em weakly} spin-dependent. I believe that this weak
spin-dependence is an additional condition for stable two-dimensional
high-$T_c$ superconducting states. This question requires further theoretical
considerations and further examinations of the band structures of high-$T_c$
superconductors.

\acknowledgements{%
  I am indebted to Ove Jepsen for providing me with all the data I needed to
  determine the symmetry of the Bloch functions in the band structure of
  YBa$_2$Cu$_3$O$_7$. I thank Ernst Helmut Brandt for critical and helpful
  comments on the manuscript.}

\end{document}